\documentstyle[preprint,aps,epsfig]{revtex}
\def\beq{\begin{equation}}
\def\eeq{\end{equation}}
\def\be{\begin{eqnarray}}
\def\ee{\end{eqnarray}}
\def\ci{\cite}
\def\bi{\bibitem}

\def\magq{|{\bf q}|}
\def\vecq{{\bf q}}
\def\primo{^\prime}
\def\shat{{\widehat S}({\bf q},t)}  
\def\bm #1{ \bbox{#1} }    
\begin{document}   

\draft

\title{Final-state interactions in the response of nuclear matter}

\author
{M. Petraki,$^1$  E. Mavrommatis,$^1$ O. Benhar,$^2$ J.~W. Clark,$^3$ 
A. Fabrocini,$^4$ and S. Fantoni$^5$}

\address
{$^1$Physics Department, Division of Nuclear and Particle Physics\\
University of Athens, GR-15771 Athens, Greece\\
$^2$INFN, Sezione di Roma 1, I-00185 Roma, Italy\\
$^3$Physics Department, Washington University, St. Louis MO 633130, USA \\
$^4$Department of Physics "E. Fermi", University of Pisa, 
 and INFN, Sezione di Pisa, I-56100 Pisa, Italy \\
$^5$International School for Advanced Studies (SISSA), I-30014 Trieste, 
Italy\\ }

\date{\today}

\maketitle

\begin{abstract}

Final-state interactions in the response of a many-body system to an 
external probe delivering large momentum are normally described 
using the eikonal approximation, for the trajectory of the struck particle,  
and the frozen approximation, for the positions of the spectators. We 
propose a generalization of this scheme, in which the initial momentum of 
the struck particle is explicitly taken into account. Numerical
calculations of the nuclear matter response at 1 $< |{\bf q}| <$ 2 GeV/c
show that the inclusion of this momentum dependence leads to a sizable effect
in the low energy tail. Possible implications for the analysis of 
existing electron-nucleus scattering data are discussed. 

\end{abstract}

\pacs{PACS numbers: 13.6.Le, 25.30Fj, 25.30Rw}


\section{Introduction}   

Final state interactions (FSI) of fast nucleons produced in electron-nucleus 
scattering at large momentum transfer have long been known to 
exert a significant effect on the coincidence $(e,e\primo p)$ cross section; 
moreover, they provide most of the strength observed in the low-energy loss 
tail of the inclusive $(e,e\primo)$ cross section (see, e.g., Ref.\ci{book}).
The main effect of FSI is a damping of the motion of the struck particle,
which can be qualitatively described in terms of the imaginary part
of the nuclear optical potential.  However, since nucleons in nuclei 
are strongly correlated, one must improve upon the optical potential 
approach rooted in a simple mean-field description of nuclear dynamics,
if one hopes to develop a fully quantitative treatment of FSI. 
It is very important to realize that nucleon-nucleon ($NN$) correlations, 
leading to large density fluctuations and to the appearance of 
high momentum components in the nuclear wave function, 
strongly affect both the initial and the final state in electron-nucleus
scattering and must be consistently taken into account.

A theoretical description of $(e,e\primo)$ processes including 
correlation effects has been developed in
Ref.\cite{Benhar91} (hereafter referred to as I) and successfully 
employed to analyze $(e,e\primo)$ data for momentum transfer in 
the range 1 GeV/c $< \magq <$ 2 GeV/c \cite{Benhar91,Benhar93,Benhar94}.
A similar approach, formulated in analogy with the theoretical
treatment of FSI in neutron scattering from quantum liquids, has been 
proposed in Ref.\ci{silcla}.

The treatment of FSI discussed in I, commonly referred to as Correlated
Glauber Approximation (CGA), rests on the assumptions that 
(i) the struck nucleon moves along a straight line with constant velocity 
(eikonal approximation), and (ii) the spectator nucleons are seen by the 
fast struck particle as a collection of fixed scattering centers 
(frozen approximation). 
The resulting inclusive cross section can be written as
a convolution integral, involving the cross section evaluated within the 
Plane-Wave Impulse Approximation (PWIA), i.e., evaluated in the absence 
of FSI, and a 
folding function embodying FSI effects.  The CGA entails
the same set of approximations as the Glauber theory 
of high-energy proton scattering off nuclei \ci{Glauber}, which has been 
successfully applied for over forty years. 

The results of I show that FSI produce a huge enhancement of the
inclusive cross section in the region of 
$\omega << \omega_{QE}$, where $\omega_{QE}$ is the energy transfer 
corresponding to elastic scattering off an isolated stationary nucleon. 
While this enhancement brings theory and experiment into agreement over a 
broad range in $\omega$, the calculated cross section substantially 
overestimates the data in the extreme low-$\omega$ tail (roughly corresponding 
to values of the Bjorken scaling variable $x=(Q^2/2m\omega)>$ 2, 
where $Q^2=\magq^2-\omega^2$ and $m$ denotes the nucleon mass). 
In order to reproduce the tail of the measured cross sections, the imaginary 
part of the free-space $NN$ scattering amplitude, which determines the shape of 
the CGA folding function, must be modified in such a way as to reduce
the effect of FSI. 

As pointed out in I, $NN$ scattering in the nuclear medium may in principle 
differ markedly from scattering in free space.  For example, Pauli 
blocking and dispersive corrections are known to be important at moderate 
energies \ci{papi}. However, their effects on the calculated cross sections 
have been 
found to be small in the kinematical region spanned by the data analyzed in
Refs.\ci{Benhar91,Benhar93,Benhar94}.  Corrections to the $NN$
amplitude associated with the extrapolation to off-shell energies are also
expected to be small\ci{bl}.

A different type of modification of the $NN$ cross section, originating from 
the internal structure of the nucleon, may play a more significant role. 
It has been suggested \ci{CT1,CT2} that
elastic scattering on a nucleon at high momentum transfer can only 
occur if the nucleon is found in the Fock state having the lowest number of 
constituents, so that the momentum transfer can be most effectively shared 
among them.
Within this picture, a nucleon is in a very compact configuration
after absorbing a large momentum ${\bf q}$.  It then travels through 
nuclear matter
experiencing very little FSI, until its standard size is recovered
on a characteristic time scale that increases with $|{\bf q}|$. 
In the limit of infinite momentum transfer, 
FSI are totally suppressed, and the nuclear medium is said to exhibit 
{\it color transparency}.

The results of the calculations of Refs.\ci{Benhar91,Benhar93,Benhar94}
show that inclusion of the effects of color transparency according to 
the model of 
Ref.\ci{CT}, {\it with no adjustable parameters}, greatly improves the
agreement between theory and data, yielding a satisfactory description of 
the low-energy loss tail of the nuclear inclusive cross sections 
for $Q^2>1.5$ (GeV/c)$^2$.     

To firmly establish the occurrence of color transparency in $(e,e\primo)$ 
processes, the accuracy of the approximations underlying the CGA must be 
carefully investigated and either validated or transcended.
In this paper, we introduce a treatment of FSI which improves upon the CGA,  
in that it allows one to take account of the initial momentum of 
the struck nucleon.  Within this approach, the response can no longer 
be written as a simple convolution integral.  However, it can still 
be expressed in terms of the spectral function and a generalized folding 
function, in a form displaying explicit dependence on both the initial 
and final momenta of the struck particle.

The theoretical description of nuclear-matter response is discussed
in Section II, where we outline the development of a systematic scheme 
that improves upon the PWIA and includes FSI effects.  The details
of the many-body calculation of the generalized folding function within 
the Fermi-HyperNetted Chain (FHNC) approach are traced in Section III.
The ensuing numerical results are presented and analyzed in Section IV,
with particular attention to the 
generalized folding functions and the nuclear matter $y$-scaling functions
at different values of $|{\bf q}|$.  The final section summarizes 
our findings and states our conclusions.


\section{Nuclear-matter response}

\subsection{Plane-Wave Impulse Approximation}        

The analysis carried out in Refs.\ci{Benhar91,Benhar93,Benhar94} required 
a full calculation of the nuclear cross section, including the electromagnetic 
vertex, as well as the use of spectral functions adapted to finite targets 
(such as those obtained within a local-density approximation \ci{Benhar94}).
In addition, since the typical momentum transfers lie in the 1--2 GeV/c
range, consistent use of relativistic kinematics was essential.

In this paper, we will avoid these complications and focus on the 
nonrelativistic response of infinite nuclear matter to a scalar 
probe, defined by
\be
\nonumber
S({\bf q},\omega) & = & \frac{1}{A} \int \frac{dt}{2\pi}  e^{i\omega t} 
\langle 0 | \rho^\dagger_{\vecq}(t)\rho_{\vecq}(0) | 0 \rangle \\
& = & \frac{1}{A} \int \frac{dt}{2\pi}\ e^{i(\omega+E_0)t} 
\langle 0 | \rho^\dagger_{\vecq} e^{-iHt} \rho_{\vecq} 
| 0 \rangle \ .
\label{resp:def}
\ee
Here $H$ and $|0\rangle$ denote the nuclear Hamiltonian and the
corresponding ground state satisfying the Schr\"odinger equation 
$H|0\rangle = E_0 |0\rangle$.  The time-dependent density fluctuation 
operator $\rho_{\vecq}(t)$ is constructed as
\beq
\rho_{\vecq}(t) = e^{iHt} \rho_{\vecq} e^{-iHt}
 = e^{iHt} \sum_{{\bf k}} a^\dagger_{{\bf k}+\vecq}a_{{\bf k}}\, e^{-iHt}\, ,
\eeq
where $a^\dagger_{{\bf k}}$ and $a_{{\bf k}}$ are nucleon creation and 
annihilation 
operators, respectively.  Note that the definition given 
in Eq.~(\ref{resp:def}) can 
be readily generalized to describe the electromagnetic response
by replacing $\rho_{\vecq}$ with the appropriate current operator.

Retaining only the incoherent contribution to the response, which is known 
to be dominant at large $\magq$, Eq.~(\ref{resp:def}) can be rewritten 
in the form
\beq
S(\vecq,\omega) = \int \frac{dt}{2\pi} e^{i(\omega+E_0)t} 
\shat \, ,
\label{sqw:ft}
\eeq
with
\beq
\shat = \int dR \, dR^\prime\, \Psi_{0}^{*}(R\primo) 
e^{-i{\vecq}\cdot{\bf r}\primo_1} 
\langle R\primo | e^{-iHt} | R \rangle  
e^{i{\bf q}\cdot{\bf r}_1}\Psi_{0}(R)\ ,
\label{incoh:resp}
\eeq
where $\{R\}=\{ {\bf r}_1,{\bf r}_2,...,{\bf r}_A \}$ specifies
the spatial configuration of the $A$-nucleon system,   
$\Psi_{0}(R)=\langle R | 0 \rangle$ is its ground-state wave function, 
and the propagator 
$U_A(R,R^\prime;t)=\langle R\primo |  e^{-iHt} | R \rangle$
represents the amplitude for the system to evolve from configuration ${R}$ 
to configuration ${R\primo}$ during the time $t$.  The wave function 
$\Psi_0(R)$ can in principle be evaluated within nonrelativistic
nuclear many-body theory.  On the other hand, the nonrelativistic approach 
cannot be used to obtain the $A$-particle propagator $U_A(R,R^\prime;t)$, 
since -- in the kinematical regime under consideration -- the struck 
nucleon typically carries a momentum larger than the nucleon mass.  
In view of the fact that a fully realistic and consistent calculation 
of $U_A(R,R^\prime;t)$ remains intractable, one must resort to
simplifying assumptions.

A systematic approximation scheme can be developed by first decomposing
the Hamiltonian according to 
\beq
H = H_{A-1} + T_{1} + H_{I}\, ,
\label{split:H}
\eeq
where $H_{{\rm A}-1}$ is the nonrelativistic Hamiltonian of the
{\it fully interacting} $(A-1)$-particle spectator system and $T_1$
denotes the Hamiltonian describing a {\it free} nucleon. The term 
\beq
H_{I} = \sum_{j=2}^A v_{1j}\, ,
\eeq
where $v_{ij}$ is the $NN$ potential, accounts for the interactions between the
struck particle and the spectators.       

The
PWIA amounts to setting $H_{I}=0$ in Eq.~(\ref{split:H}), thus disregarding FSI
altogether. The resulting $A$-particle propagator factorizes into the product
of the interacting $(A-1)$-particle propagator and the free-space one-body 
propagator describing the struck nucleon:
\beq
U_{\rm PWIA}(R,R^\prime;t) = U_{A-1}(\widetilde{R},\widetilde{R}^\prime;t) 
U_0({\bf r}_1,{\bf r}^\prime_1;t)\, ,
\label{PWIA:prop}
\eeq
where $\{\widetilde{R}\}=\{{\bf r}_2,...,{\bf r}_A\}$ specifies the
configuration of the spectator system. Eq~.(\ref{PWIA:prop}) clearly shows
that within the PWIA, nuclear dynamics only appears through $U_{A-1}$, 
while the 
treatment of the relativistic motion of the struck nucleon reduces to a 
trivial kinematic problem.      
                                                   
We next express the PWIA to $S({\bf q},t)$ (and thereby the response)
in terms of the nucleon spectral function $P({\bf k},E)$, which by definition
gives the probability of removing a nucleon with momentum ${\bf k}$
from the nuclear ground state, leaving the residual system
with excitation energy $E$.  Introducing spectral representations for 
both $U_{A-1}$ and $U_0$ (see, e.g., Ref.\ci{PRL2001}), we obtain
\beq
\widehat{S}_{\rm PWIA}({\bf q},t) = \int \frac{d^3p}{(2\pi)^3} \int dE \,
P({\bf p}-{\bf q},E) {\rm e}^{-i(E-E_0+E_p)t}\, ,
\label{PWIA:shat}
\eeq
which leads to the familiar result \ci{PRL2001}
\beq
S_{\rm PWIA}({\bf q},\omega) =
\int \frac{d^3p}{(2\pi)^3}  \int dE\,
 P({\bf p}-{\bf q},E) \delta(\omega - E - E_p)\, ,
\eeq
where $E_p=|{\bf p}|^2/2m$ denotes the kinetic energy of a nucleon carrying 
momentum ${\bf p}$.

\subsection{Inclusion of Final-State Effects}

In order to improve upon the PWIA, one needs to devise
a set of sensible approximations to treat the contributions to the
$A$-particle propagator coming from the FSI Hamiltonian $H_I$. As a 
first step, we make the replacement
\beq
e^{-i(H_{A-1} + T_1 + H_I)t} \rightarrow 
e^{-iH_{A-1}t} e^{-i(T_1 + H_I)t}\, ,
\eeq
which essentially amounts to assuming that the internal dynamics of
the spectator system and its FSI with the struck particle do not
affect one another, and can therefore be completely decoupled.
Within this picture, the spectator system evolves during the
time $t$ as if there were
no struck particle moving around, while the fast struck particle
sees the spectator system as if it were {\it frozen} at time $t=0$. 

Under this assumption, which implies that the
configuration of the spectator system does not change due to
interactions with the fast struck nucleon, we can use
completeness of the $(A-1)$-particle position 
eigenstates to rewrite the propagator in the simple factorized form
\begin{eqnarray}
\nonumber
U_A(R,R^\prime;t) & = & \int d\widetilde{R}^{\prime\prime}
\langle \widetilde{R}^\prime | {\rm e}^{-iH_{A-1}t} | 
\widetilde{R}^{\prime\prime} \rangle
\langle {\bf r}\primo_1, \widetilde{R}^{\prime\prime} | 
e^{-i(T_1 + H_I)t}| {\bf r}_1,\widetilde{R} \rangle 
\delta^{3(A-1)}(\widetilde{R}-\widetilde{R}^{\prime\prime}) \\
\nonumber
& = & U_{A-1}(\widetilde{R},\widetilde{R}\primo;t)
\langle {\bf r}\primo_1, \widetilde{R} | 
e^{-i(T_1 + H_I)t}| {\bf r}_1, \widetilde{R} \rangle \\
& = & U_{A-1}(\widetilde{R},\widetilde{R}\primo;t)\, 
U_1( {\bf r}_1 \widetilde{R}, {\bf r}\primo_1 \widetilde{R}; t)\, .
\label{FSI:prop}
\end{eqnarray}
Evaluation of $U_{1}({\bf r}_1 \widetilde{R},
{\bf r}\primo_1 \widetilde{R};t)$ in general requires a 
functional integration over the set of 
trajectories ${\bf r}_1(\tau)$ satisfying the boundary
conditions ${\bf r}_1(0)={\bf r}_1$ and ${\bf r}_1(t)={\bf r}^{\prime}_1$
(see, e.g., Ref.\ci{Negele}). 
However, for large momenta of the struck nucleon, the evaluation can be
drastically simplified by invoking the {\it eikonal approximation}, i.e. 
by
assuming that 
the particle moves along a straight trajectory with constant velocity
${\bf v}=({\bf r}\primo_1 - {\bf r}_1)/t$, so that 
${\bf r}_1(\tau) = {\bf r}_1 + {\bf v}\tau\ $.
Within this approximation, the propagator $U_{1}({\bf r}_1 \widetilde{R},
{\bf r}\primo_1 \widetilde{R};t)$ takes the factorized form 
\cite{Carraro91}
\beq
U_{1}({\bf r}_1 \widetilde{R}, {\bf r}\primo_1 \widetilde{R};t) =
U_0({\bf r}_1,{\bf r}^\prime_1;t) U_{\bf p}({\bf r}_1,\widetilde{R};t) \, ,
\eeq  
where the eikonal propagator $U_{\bf p}({\bf r}_1,\widetilde{R};t)$ is 
given by
\beq
U_{\bf p}({\bf r}_1,\widetilde{R};t) = \exp { \left[ {- i \int_0^t d\tau
\sum_{j=2}^A  v({\bf r}_1 + {\bf v}\tau - {\bf r}_j)}  \right] }\, ,
\label{eikonal:prop}
\eeq
$v$ being the $NN$ potential and $U_0({\bf r}_1,{\bf r}^\prime_1;t)$ the 
free-space nucleon propagator. 

Expanding the exponential appearing in the right-hand side of 
Eq.~({\ref{eikonal:prop}), one obtains a series whose terms are associated with
processes involving an increasing number of interactions between the struck 
nucleon and the spectators. The terms corresponding to repeated
interactions with the same spectator can be summed up to all orders 
by replacing the bare $NN$ interaction
$v$ with the coordinate space $t$-matrix $\Gamma_{\bf q}({\bf r})$, 
which is related to the $NN$ scattering amplitude $f_{\bf q}({\bf k})$ 
at incident momentum ${\bf q}$ and momentum transfer 
${\bf k}$ through
\beq
\Gamma_{\bf q}({\bf r}) = - \frac{2 \pi}{m} \int \frac{d^3k}{(2 \pi)^3}
e^{i{\bf k}\cdot{\bf r}} f_{\bf q}({\bf k})\, .
\label{def:Gamma}
\eeq

Using the above results together with spectral representations of both 
$U_0({\bf r}_1,{\bf r}^\prime_1;t)$ and 
$U_{A-1}(\widetilde{R},\widetilde{R}^{\prime};t)$, 
the response can finally be expressed as
\beq
S({\bf q},\omega)  = \int  \frac{dt}{2\pi} 
e^{i(\omega + E_0) t} \int  \frac{d^3p}{(2\pi)^3} e^{-iE_pt} 
 \sum_n e^{-iE_nt} M_{0n}^*({\bf p} - {\bf q})
\widetilde{M}_{0n}({\bf p},{\bf q};t) \, ,
\label{resp:over}
\eeq
where the sum extends over the $(A-1)$-particle states satisfying the
Schr\"odinger equations $H_{A-1}|n\rangle=E_n|n\rangle$.  We have
introduced the definitions 
\beq
M_{0n}({\bf k}) = \int dR\, e^{i{\bf k}\cdot{\bf r}_1}
\Psi_{0}^{*}(R) \Phi_{n}(\widetilde{R})\, ,
\label{mon:def}   
\eeq
with $\Phi_{n}(\widetilde{R}) = \langle \widetilde{R} | n \rangle$, and
\beq
\widetilde{M}_{0n}({\bf p},{\bf q};t) =  \int  dR\, 
e^{-i({\bf p} - {\bf q})\cdot{\bf r}_1}
\Phi_{n}^{*}(\widetilde{R})\Psi_{0}(R)
U_{\bf p}({\bf r}_1,\widetilde{R};t)\, .
\label{mtilde:def}
\eeq
In the limit $U_{\bf p}({\bf r}_1,\widetilde{R};t) \rightarrow 1$, we have
$\widetilde{M}_{0n}({\bf p},{\bf q};t) \rightarrow M_{0n}({\bf p}-{\bf q})$,
and the response $S({\bf q},\omega)$ given by Eq.~(\ref{resp:over}) 
reduces to the PWIA result.
           
It is to be emphasized that the calculation of the response according 
to Eq.~(\ref{resp:over}) involves only two approximations: (i) the frozen 
approximation for the configuration of the spectator 
system and (ii) the eikonal approximation for the trajectory of the struck 
particle. 
Explicit calculation of the relevant 
$\widetilde{M}_{0n}({\bf p},{\bf q};t)$ integrals
within nonrelativistic many-body theory appears to be feasible, at least for
few-nucleon systems and infinite nuclear matter. 
However, to establish a clear connection with 
the PWIA picture, it is useful to devise approximations that permit
$S({\bf q},\omega)$ of Eq.~(\ref{resp:over}) to be expressed in terms 
of either the spectral function  $P({\bf k},E)$ or 
the PWIA response $S_{\rm PWIA}({\bf q},\omega)$.           

The spectral function is defined as
\beq
P({\bf k},E) = \sum_n | {M}_{0n}({\bf k}) |^2 \delta(E+E_0-E_n)\, .
\eeq
To recover this quantity within the formula (15), 
the integrals $\widetilde{M}_{0n}$ of Eq.~(\ref{mtilde:def}) 
are required to take the form
\beq
\widetilde{M}_{0n}({\bf p},{\bf q};t) = {M}_{0n}({\bf p}-{\bf q})\,
 {\cal U}({\bf p},{\bf q};t)\, ,
\label{mtilup:def}
\eeq
where the function ${\cal U}$ is to be independent of the state of the
spectator system, labeled by the index $n$.
An even more drastic simplification is achieved upon requiring that the time 
dependence of $\widetilde{M}_{0n}({\bf p},{\bf q};t)$ be factorizable 
according to
\beq
\widetilde{M}_{0n}({\bf p},{\bf q};t) = 
M_{0n}({\bf p}-{\bf q}) {\overline U}_{\bf q}(t)\, ,
\label{fact:t}
\eeq
i.e.\ upon assuming that the function ${\cal U}$ defined by
Eq.~(\ref{mtilup:def}) does not depend upon ${\bf p}$, which in turn 
corresponds to making the approximation ${\bf p} \simeq {\bf q}$ in 
${\cal U}$.    

Substitution of Eq.~(\ref{fact:t}) into Eq.~(\ref{resp:over}) allows one to 
rewrite the response as a convolution integral,
\beq
S({\bf q},\omega) = \int  d\omega^\prime S_{\rm PWIA}({\bf q},\omega^\prime) 
F_{\bf q}(\omega-\omega^\prime)\, ,
\label{convolution}
\eeq
the folding function $F_{\bf q}(\omega)$ being given by
\beq
F_{\bf q}(\omega) = \int  \frac{dt}{2\pi}  e^{i \omega t}
{\overline U}_{\bf q}(t)\, .
\eeq
To obtain the function ${\overline U}_{\bf q}(t)$ embodying all FSI 
effects ({\it N.B.} the PWIA can be regained by setting
${\overline U}_{\bf q}(t) \equiv 1$, i.e., $F(\omega) = \delta(\omega)$), 
one makes the replacement 
\beq
\sum_{j=2}^A \Gamma_{\bf q}({\bf r}_1 + {\bf v}\tau - {\bf r}_j) 
\rightarrow  \frac{ \int   dR\, |\Psi_0(R)|^2 \sum_{j=2}^A
\Gamma_{\bf q}({\bf r}_1 + {\bf v}\tau - {\bf r}_j) }
{ \int dR\, |\Psi_0(R)|^2 } = {\overline V}_{\bf q}(\tau)\, , 
\label{average}
\eeq
which amounts to averaging FSI with the ground-state configuration
weights.
In infinite nuclear matter at uniform density $\rho$, the average involved
in Eq.~(\ref{average}) takes the simple form
\beq
{\overline V}_{\bf q}(\tau) = \rho \int d^3r\, g(r)
\Gamma_{\bf q}({\bf r} + {\bf v}\tau)\, ,
\label{Vbar:nm}
\eeq
where the radial distribution function $g(r)$ measures the probability of
finding two nucleons separated by a distance $r=|{\bf r}|$.
Keeping only the contributions associated with the imaginary part of the 
$NN$ amplitude, which is known to be dominant 
at large incident momentum, we can finally write the eikonal propagator
as
\beq
{\overline U}_{\bf q}(t) = \exp { \int d\tau \,
{\rm Im}  {\overline V}_{\bf q}(\tau) }\, .
\label{averaged:prop}
\eeq

The approach developed in I and employed in 
Refs.\cite{Benhar91,Benhar93,Benhar94} is based on the assumptions underlying
Eqs.~(\ref{fact:t})--(\ref{averaged:prop}).  A different approximation scheme 
leading to the factorization of $\widetilde{M}_{0n}$
can be obtained by inserting into Eq.~(\ref{mtilde:def}) the identity
\beq
\int d\tilde{R}\primo d^3r\primo_1
\delta({\bf r}_1-{\bf r}\primo_1)\, \delta^{3(A-1)} (R - R\primo) =
\sum_N \int dR\primo\, \Psi^{*}_N(R\primo) \Psi_N (R)\, ,
\label{deltaR}
\eeq 
where the sum includes a complete set of eigenstates of the
$A$-particle Hamiltonian $H$.  This procedure is not unique,
because the $\delta$-function insertion allows for different
assignments of the arguments of the functions entering the right-hand side of
Eq.~(\ref{mtilde:def}), leading to different but ultimately equivalent
representations.  Equating two such representations, we have
$$
  \sum_{N} \int d\tilde{R}\, d^3r_1\, e^{-i{\bf k}\cdot{\bf r}_1}
  \Phi^*_n(\widetilde{R})\Psi_N({\bf r}_1,\widetilde{R})
 \int d\widetilde{R}\primo  d^3r\primo_1 \,
 e^{i{\bf k}\cdot({\bf r}_1 - {\bf r}\primo_1)}
\Psi_0({\bf r}\primo_1,\widetilde{R}\primo)
\Psi_N^*({\bf r}_1,\widetilde{R}\primo) 
U_{\bf p}({\bf r}\primo_1,\widetilde{R}\primo;t)
$$
\beq
  =  \sum_{N} \int d\widetilde{R}\,  d^3r_1 \,
 e^{-i{\bf k}\cdot{\bf r}_1}
  \Phi^*_n(\widetilde{R})\Psi_N(R) U_{\bf p}({\bf r}_1, \widetilde{R};t)
   \int d\widetilde{R}\primo   d^3r\primo_1 \,
 e^{i{\bf k}\cdot({\bf r}_1 - {\bf r}\primo_1)}
\Psi_0({\bf r}\primo_1,\widetilde{R}\primo) 
\Psi_N^*({\bf r}_1,\widetilde{R}\primo)\, ,
\label{eq:5}
\eeq
with ${\bf k}={\bf p}-{\bf q}$.  Now imposing translation invariance, 
as appropriate for the case of infinite nuclear matter, and
retaining only the term corresponding to $|N\rangle = |0\rangle$ in
the above sums, Eq.~(\ref{eq:5}) becomes   
\beq
\widetilde{M}_{0n}({\bf p},{\bf q};t) = M_{0n}({\bf k})\ \frac 
{{\cal U}_0({\bf k},{\bf p};t)}{n({\bf k})}\, .
\label{mtilapp}
\eeq
Here $n({\bf k})$ is the nucleon momentum distribution,       
defined in terms of the spectral function as 
\beq 
n({\bf k}) = \int dE\, P({\bf k},E)\, ,
\eeq
while 
\beq
{\cal U}_0({\bf k},{\bf p};t) =  \int d^3 r_{11\primo}\, 
{\rm e}^{i{\bf k}\cdot{\bf r}_{11\primo}}
{\cal P}({\bf p},{\bf r}_{11\primo};t) \, ,
\label{def:u0}
\eeq
with ${\bf r}_{11\primo}={\bf r}_1-{\bf r}_{1\primo}$ and
\beq
{\cal P}({\bf p},{\bf r}_{11\primo};t) = \int d\widetilde{R}\,
 \Psi_{0}^{*}({\bf r}\primo_1,\widetilde{R})\Psi_{0}({\bf r}_1,
\widetilde{R}) U_{\bf p}({\bf r}_1,\widetilde{R};t)\, .
\label{def:p}
\eeq               
It can be readily seen that if $U_{\bf p}(\tilde{R},{\bf r}_1;t)\equiv 1$, 
i.e.\ if FSI are absent, the function ${\cal P}({\bf p},{\bf r}_{11\primo};t)$ 
reduces to $\rho({\bf r}_1;{\bf r}_{1\primo})/A$, where 
\beq
\rho_{1}({\bf r}_{1};{\bf r}_{1}^{\prime}) = A \int d\widetilde{R}\,
\Psi^*_0({\bf r}\primo_1,\widetilde{R})
\Psi_0({\bf r}_1,\widetilde{R})
\label{obdm}
\eeq
is the one-body density matrix, whose Fourier transform is $n({\bf k})$. 
As a consequence, we have
$\widetilde{M}_{0n}({\bf p},{\bf q};t) = M_{0n}({\bf k})$ and the PWIA is
recovered. In addition, the relation (\ref{mtilapp}) can be shown to 
hold as an equality at the lowest order of the cluster expansion.           

Substitution of Eq.~(\ref{mtilapp}) into the definition of the response 
leads to
\beq
S({\bf q},\omega) = \int  d\omega\primo \int 
\frac {d^3k}{(2 \pi)^3} {F}_{{\bf k},{\bf q}}(\omega\primo - \omega)
 \int  dE\, P({\bf k},E) \delta(\omega\primo-E-E_{|{\bf k}+{\bf q}|})\, ,
\eeq
where the generalized folding function ${F}_{{\bf k},{\bf q}}(\omega)$
is defined as
\beq
{F}_{{\bf k},{\bf q}}(\omega) =
\frac{1}{n({\bf k})}\int  \frac{dt}{2\pi} {\rm e}^{i\omega t} \,
{\cal U}_0({\bf k},{\bf k}+{\bf q};t)\, .
\label{def:genf}
\eeq                                         

\section{Many-Body Calculation of the Generalized Folding Function}

Eqs.~(\ref{def:u0}) and (\ref{def:p}) show that calculation of the 
generalized folding function of Eq.~(\ref{def:genf}) requires 
a knowledge of the partially diagonal $n$-body density matrices 
\beq
\rho_{n}({\bf r}_{1},{\bf r}_{2},\ldots,{\bf r}_{n};
{\bf r}_{1}^{\prime},{\bf r}_{2},\ldots,{\bf r}_{n})
= \frac{A!}{(A-n)!}\, 
\int d^3r_{n+1}\ldots d^3r_A \Psi^*_0({\bf r}\primo_1,\widetilde{R})
\Psi_0({\bf r}_1,\widetilde{R})
\label{def:densmat}
\eeq
of the target nucleus, for all $n\leq$A. 

The numerical calculation of $\rho_{n}$ within 
an {\it ab initio} microscopic approach involves 
prohibitive difficulties, even for the 
case of infinite nuclear matter considered here.
In view of this problem, we need to model $\rho_{n}$ in terms 
of quantities that consistently incorporate the relevant physics
and can still be reliably calculated. The results presented in this 
paper have been obtained using an approximation scheme 
(hereafter referred to as the {\it hole approximation}), 
in which $\rho_{n}$ is written in terms of the one-body density matrix
of Eq.~(\ref{obdm}) and the half-diagonal two-body density matrix
\beq
\rho_{2}({\bf r}_{1},{\bf r}_{2};{\bf r}_{1}^{\prime},{\bf r}_{2}) 
= A(A-1) \int d^3r_3 \ldots d^3r_A
\Psi^*_0({\bf r}\primo_1,\widetilde{R})
\Psi_0({\bf r}_1,\widetilde{R})\, .
\label{2hdm}
\eeq
The explicit formula for the resulting $n$-body density matrix, 
\begin{equation}
\rho_{n}^{\rm HA}({\bf r}_{1},{\bf r}_{2},\ldots ,{\bf r}_{n};
{\bf r}_{1}^{\prime},{\bf r}_{2},\ldots,{\bf r}_{n})
 =  
\frac{1}{[\rho_{1}({\bf r}_{1};{\bf r}_{1}^{\prime})]^{n-2}}
\prod_{i=2}^{n}\rho_{2}({\bf r}_{1},{\bf r}_{i};
{\bf r}_{1}^{\prime},{\bf r}_{i}) \, ,
\label{eq:HAI}
\end{equation}
shows that the hole approximation represents
the multinucleon spatial correlations involved in the definition of 
$\rho_{n}$ as a superposition of two-nucleon correlations. 

Among other expressions that can be constructed from the same
building blocks, Eq.~(\ref{eq:HAI}) was chosen primarily because
it fulfills some basic 
properties of the exact density matrices. 
In particular, $\rho_{n}^{\rm HA}$, which is obviously real, satisfies exactly 
the asymptotic factorization requirement
\beq
\lim_{|{\bf r}_n| \rightarrow \infty}
\rho_{n}({\bf r}_{1},{\bf r}_{2},\ldots ,{\bf r}_{n};
{\bf r}_{1}^{\prime},{\bf r}_{2},\ldots,{\bf r}_{n})
= \rho \rho_{n-1}({\bf r}_{1},{\bf r}_{2},\ldots ,{\bf r}_{n-1};
{\bf r}_{1}^{\prime},{\bf r}_{2},\ldots,{\bf r}_{n-1})\, ,
\eeq
while violating, although not severely, the sequential relation 
\begin{eqnarray}
\nonumber
 & & \int d^3r_n\, \rho_{n}({\bf r}_{1},{\bf r}_{2},\ldots ,{\bf r}_{n};
{\bf r}_{1}^{\prime},{\bf r}_{2},\ldots,{\bf r}_{n}) \\
 & & \ \ \ \ \ \ \ \ \ \ \ \ \ \  = 
\left[A-(n-1)\right] 
\rho_{n-1}({\bf r}_{1},{\bf r}_{2},\ldots ,{\bf r}_{n-1};
{\bf r}_{1}^{\prime},{\bf r}_{2},\ldots,{\bf r}_{n-1})\, .
\label{seq}
\end{eqnarray}
Within the hole approximation, Eq.~(\ref{seq}) translates into
\begin{eqnarray}
\nonumber
& & \int d^3r_n\, \rho_{n}^{\rm HA}({\bf r}_{1},{\bf r}_{2},\ldots,{\bf r}_{n};
{\bf r}_{1}^{\prime},{\bf r}_{2},\ldots,{\bf r}_{n}) \\
& & \ \ \ \ \ \ \ \ \ \ \ \ \ \ \ = 
(A-1) \rho_{n-1}^{\rm HA}({\bf r}_{1},{\bf r}_{2},\ldots,{\bf r}_{n-1};
{\bf r}_{1}^{\prime},{\bf r}_{2},\ldots,{\bf r}_{n-1}) \\ 
& & \ \ \ \ \ \ \ \ \ \ \ \ \ \ \ = 
[A-(n-1)]\rho_{n-1}^{\rm HA} + O(n-2)\, .
\end{eqnarray}

In addition to $\rho_2$, calculation of the generalized folding function  
in the hole approximation calls for a knowledge of the 
imaginary part of the quantity $\Gamma_{\bf q}$ of 
Eq.~(\ref{def:Gamma}), i.e.\ of the imaginary part 
of the $NN$ scattering amplitude $f_{\bf q}$. In this work, 
we have employed the simple parametrization 
originally proposed in Ref.\cite{Bassel-Wilkin67}, namely
\beq
{\rm Im} f_{\bf q}({\bf k}) = \frac{|{\bf q}|}{4 \pi}\, \sigma_{NN}
\exp{ \left( -\beta^2 |{\bf k}|^2 \right) }\, .
\eeq
Numerical values of the total $NN$ cross section $\sigma_{NN}$ and the
slope parameter $\beta$ resulting from fits to $NN$ scattering data
are given in Refs.\cite{Dobrovolsky-et-al83} and \cite{Silverman-et-al89}.

Using the hole approximation (\ref{eq:HAI}) and together with the 
parametrization (42) of ${\rm Im} \Gamma_{\bf q}$, we can finally 
assemble the working expression
\beq
{F}_{{\bf k},{\bf q}}(\omega) = \frac{1}{n({\bf k})} 2 {\rm Re}\, 
 \frac{1}{v} \int_{0}^{\infty}  \frac{dz}{2\pi}  
\exp{ \left( i\omega \frac{z}{v} \right) }  \int d{\bf r}_{11^\prime}\, 
e^{i{\bf k}\cdot{\bf r}_{11^\prime}} E_{\bf q}(z,{\bf r}_{11^\prime})
\label{eq:fol5}
\eeq
for the generalized folding function, with
\beq
E_{\bf q}(z,{\bf r}_{11^\prime}) = 
\exp{ \left[ -\frac{\sigma_{NN}}{16\pi \beta^{2}}
J_{\bf q}(z,{\bf r}_{11^\prime}) \right] }
\label{eq:fol5b}
\eeq
and 
\begin{eqnarray}
\nonumber
J_{\bf q}(z,{\bf r}_{11^\prime}) & = & \frac{1}{\rho_{1}(r_{11^\prime})}
\int_{0}^{\infty} r^2 dr \int_{-1}^{1}
d\cos \theta\ \rho_{2}(r_{11^\prime},r,r^\prime) \\
& \qquad & \times
\exp{ \left[ -\frac{r^{2}\sin^{2}\theta}{4\beta^{2}} \right] }
\left[ {\rm erf} \left( \frac{z + r\cos\theta}{2 \beta} \right) 
     - {\rm erf} \left( \frac{r\cos\theta}{2 \beta} \right)   \right] \, .
\label{eq:fol5c}
\end{eqnarray}
In the above equations, $\rho_{1}(r_{11^\prime})$ and 
$\rho_{2}(r_{11^\prime},r,r^\prime)$, with 
$r = |{\bf r}| = |{\bf r}_1 - {\bf r}_2|$ and 
$r^\prime = |{\bf r}^\prime| = |{\bf r}_{1}^{\prime} -
{\bf r}_2|$, are the nuclear-matter 
one-body and half-diagonal two-body density matrices, respectively, 
$v$ is the speed of the struck particle, the $z$ axis is 
chosen along the direction of ${\bf q}$, and $\theta$ is the angle between
${\bf r}$ and ${\bf q}$.  It can be shown that the 
${\bf k}$-independent CGA folding function of Ref.\cite{Benhar91} 
can be recaptured from Eqs.~(43)--(45) by replacing
$\rho_{2}({\bf r}_{1},{\bf r}_{2};{\bf r}_{1}^{\prime},{\bf r}_{2})$ 
with its fully diagonal part 
$\rho_{2}({\bf r}_{1},{\bf r}_{2};{\bf r}_{1},{\bf r}_{2})$.  
The numerical calculation of ${F}_{{\bf k},{\bf q}}(\omega)$ has
been performed by applying (i) the scheme proposed in Ref.\cite{Flynn-et-al84} 
to obtain $n(k)$ and $\rho_{1}(r_{11^{'}})$ and (ii) the formalism 
developed in Ref.\cite{Petraki-et-al00} to evaluate the nuclear-matter 
half-diagonal two-body density matrix.  Both approaches use 
correlated many-body wave functions and FHNC integral equations to sum up 
selected cluster contributions to all orders in the relevant expansions.

In particular, $\rho_{2}({\bf r}_{1},{\bf r}_{2};
{\bf r}_{1}^{\prime},{\bf r}_{2})$ has been approximated by its 
leading $dd$ part in the FHNC formalism according to 
\ci{Petraki-et-al00}; thus we set
\begin{equation}
\rho_{2}({\bf r}_{1},{\bf r}_{2};{\bf r}_{1}^{\prime},{\bf r}_{2})
=
\rho_{1}(r_{11^\prime})
g_{Qdd}(r) g_{Qdd}(r^\prime)\, ,
\end{equation}
where $g_{Qdd}(r)$ consists of the  $dd$ nodal and non-nodal
components of the FHNC expression for 
$\rho_{1}({\bf r}_{1};{\bf r}_{1}^{\prime})$.  (The notation $dd$ 
refers to the topological classification of the diagrams associated 
with the corresponding terms in the cluster expansion. 
See Ref.\cite{Petraki-et-al00} for details.)

Due to its weak dependence upon $\cos\phi = 
{\widehat {\bf r}}\cdot{\widehat {\bf r}}_{11^\prime}$, 
the quantity $g_{Qdd}(r^\prime)$ is approximated by its angular
average, according to 
\begin{equation}
g_{Qdd}(r_1^\prime) \rightarrow 
\overline{g}_{Qdd}(r^\prime) 
\equiv \frac{1}{2}
\int_{-1}^{1} d(\cos\phi) \, g_{Qdd}\left[ (r^{2}+r_{11^{'}}^{2}
-2rr_{11^{'}}\cos\phi)^{1/2} \right]\, . 
\label{eq:approxg}
\end{equation}
The resulting expression for the generalized folding function is
\beq
{F}_{{\bf k},{\bf q}}(\omega) = \frac{4}{n({\bf k})}
\int_{0}^{\infty} \frac{dz}{v} \cos \left( \omega \frac{z}{v} \right)
\int_{0}^{\infty}\ r^2_{11^\prime} dr_{11^\prime} \,
j_0(kr_{11^\prime}) \rho_{1}(r_{11^\prime}) 
{\bar E}_{\bf q}(z,r_{11^\prime}) \, ,
\label{eq:fol8}
\eeq
where $j_0(x)$ is the zeroth-order spherical Bessel function and  
\beq
{\bar E}_{\bf q}(z,{\bf r}_{11^\prime})  =  
\exp{ \left[ -\frac{\sigma_{NN}}{16\pi \beta^{2}} 
{\bar J}_{\bf q}(z,{\bf r}_{11^\prime}) \right] }
\label{eq:fol5bn}
\eeq
with 
\begin{eqnarray}
\nonumber
{\bar J}_{\bf q}(z,{\bf r}_{11^\prime}) & = & 
\ 2 \pi \rho \int_{0}^{\infty} r^2 dr \int_{-1}^{1}
d\cos \theta\, g_{Qdd}(r) {\overline g}_{Qdd}(r^\prime) \\
& \qquad & \times \exp{ \left[ -\frac{r^{2}\sin^{2}\theta}{4\beta^{2}} \right] }
\left[ {\rm erf} \left( \frac{z + r\cos\theta}{2 \beta} \right)
     - {\rm erf} \left( \frac{r\cos\theta}{2 \beta} \right)   \right] \, .
\label{eq:calE-last}
\end{eqnarray}

\section{Numerical Results}

We have carried out the calculation of the nonrelativistic response of 
a realistic model of nuclear matter, based on the Hamiltonian
\beq
H = \sum_{i=1}^A \frac{{\bf p}^2}{2m} + \sum_{j>i=1}^A v_{ij} 
 + \sum_{k>j>i=1}^A\ V_{ijk}\, ,
\label{nucl:ham}
\eeq
where $v_{ij}$ and $V_{ijk}$ are potentials describing two- and 
three-nucleon interactions, and on a variational ground-state wave function
\beq
\Psi_{0}(R) = {\cal S} \bigl[   \prod_{i<j} F_{ij}  \bigr]  \chi_0(R)\, .
\label{def:NMWF}
\eeq 
In this trial form, ${\cal S}$ is a symmetrization operator acting on 
the product of two-nucleon correlation operators, $F_{ij}$, and 
$\chi_0$ is the Slater determinant describing a noninteracting Fermi gas 
of nucleons with momenta ${\bf k}$ filling the Fermi sea, i.e., with
$|{\bf k}|\le k_F = (3 \pi^2 \rho/2)^{1/3}$. The operator $F_{ij}$,
which should reflect the correlation structure induced by
the nuclear Hamiltonian, has been chosen as \cite{Pan79}
\begin{eqnarray}
F_{ij} & = & f_c(r_{ij}) + f_\sigma(r_{ij}) (\bm\sigma_i\cdot\bm\sigma_j)
 + f_\tau(r_{ij}) (\bm\tau_i\cdot\bm\tau_j)
 + f_{\sigma\tau}(r_{ij}) (\bm\sigma_i\cdot\bm\sigma_j)
( \bm\tau_i\cdot\bm\tau_j)
 \nonumber \\
 &   & \ \ \ \ \ \ \ \ \ \ \ \ \ \ + \ f_t(r_{ij})\ S_{ij}
+ f_{t\tau}(r_{ij})\  S_{ij}\ (\bm\tau_i\cdot\bm\tau_j)\, .
\label{corr.op.-coord.}
\end{eqnarray}
Here $S_{ij} = 3(\bm\sigma_i \cdot {\bf r}_{ij})
(\bm\sigma_j \cdot {\bf r}_{ij})/|{\bf r}_{ij}|^2 - 
(\bm\sigma_i\cdot\bm\sigma_j)$ is the usual tensor operator, while
$f_c(r), f_\sigma(r), f_\tau(r), f_{\sigma\tau}(r), f_t(r)$, and
$f_{t\tau}(r)$ are correlation functions whose radial shapes
are determined by minimizing the expectation value of the Hamiltonian
(\ref{nucl:ham}) in the ground state described by Eq.~(\ref{def:NMWF}) 
\cite{Pan79}. 

The PWIA response has been calculated using the nucleon spectral function
of Ref.\ci{BFF89}, obtained from a nuclear Hamiltonian including
the Urbana $v_{14}$ $NN$ potential supplemented by the TNI model of the
three-body interaction \cite{Lag81}. The same ingredients entering the
calculation of the spectral function have been employed in the calculation
of the density matrices needed to obtain the generalized folding function
defined by Eqs.~(\ref{eq:fol8})--(\ref{eq:calE-last}).

The dependence of the folding function $F_{{\bf k},{\bf q}}(\omega)$
on $k \equiv |{\bf k}|$ and $\omega$ is illustrated in Figs.~1 and 2 at  
$|{\bf q}| =$ 1.31 and 1.94 GeV/c, two values representative of the range 
covered by the data analyzed in Refs.\ci{Benhar91,Benhar93,Benhar94}.
For comparison, we also show the 
$k$-independent folding functions obtained using the approach 
developed in I.
We note that in symmetric nuclear matter at its equilibrium density
$\rho=\rho_0= 0.16$ fm$^{-3}$, the Fermi momentum is $k_F=1.33$ fm$^{-1}$.
It is apparent that in the region $k<k_{F}$, the folding function
$F_{{\bf k},{\bf q}}(\omega)$ is very close to its CGA counterpart,
whereas a strong $k$-dependence is observed at $k>k_{F}$.
At $|{\bf k}| \gtrsim k_F$ the generalized folding function shrinks, and its
tail begins to oscillate, implying that these $k$ values correspond
to weaker FSI.  On the other hand, for larger momenta, well
above the Fermi level, ${F}_{{\bf k},{\bf q}}(\omega)$ gets broader again.    
The results of Figs.~\ref{GFF1} and \ref{GFF2} confirm the naive
expectation that while the averaging procedure involved in the approach
of Ref.\ci{Benhar91} is quite reasonable in the region of
$|{\bf k}| < k_F$, where the nucleon momentum distribution is nearly
constant, the momentum dependence of the folding function associated with
fast nucleons, carrying momenta larger than $k_F$, must be treated 
explicitly.                                

Figs.~\ref{FY1} and \ref{FY2} show the dependence of the nuclear 
matter $y$-scaling function
\beq
F({\bf q},y) = \frac{|{\bf q}|}{m} S({\bf q},y)
\label{def:Fy}
\eeq
on the scaling variable
\beq
y = \frac{m}{|{\bf q}|} \left( \omega - \frac{|{\bf q}|^2}{2m} \right)\, , 
\label{def:y}
\eeq
evaluated for the respective choices $|{\bf q}| =$ 1.94 and 1.31 GeV/c 
of the momentum transfer.
Using the $y$-scaling function rather than the response makes it 
easier to directly compare FSI effects at different values of $|{\bf q}|$, 
as measured by the deviation from the PWIA results at fixed $y$.

Comparison of the PWIA scaling functions (dash-dot lines) 
to the results obtained from the CGA (dashed lines) and the approach 
described in the previous sections (solid lines) clearly demonstrates that 
use of the ${\bf k}$-dependent generalized folding 
function leads to a suppression of FSI effects in the region of large negative 
$y$, corresponding to the low-energy tail of the response 
[see Eq.~(\ref{def:y})].  For example, at $y$ = 600 MeV/c the
differences between the dashed and solid lines are $\simeq$ 70 \%
and $\simeq$ 20 \% at $|{\bf q}|$ = 1.31 and 1.94 GeV/c, respectively.

The fact that the suppression of FSI appears to be more 
pronounced at the lower values of $|{\bf q}|$ indicates that a broader
CGA folding function is associated with a smaller effect of 
${\bf k}$-dependence. In fact, the larger value of the $NN$ cross section at
$|{\bf q}|$ = 1.94 GeV/c, namely $\sigma_{NN} \simeq $ 43 mb compared to
$\sigma_{NN} \simeq $ 35 mb at $|{\bf q}|$ = 1.31 GeV/c, makes the 
CGA folding function broader at the higher momentum transfer (see
Figs.\ref{GFF1} and \ref{GFF2}).
 
\section{Summary and Conclusions}

We have carried out a calculation of the nuclear matter response in which 
$NN$ correlations, which are known to play an important role
in the low-energy tail of the response, have been taken into account both 
in the initial and final states.
The effects of dynamical correlations in the initial state have been 
consistently incorporated into the PWIA calculation, based on a realistic 
spectral function obtained from an {\it ab initio} microscopic many-body 
approach \ci{BFF89}. 

The same ingredients entering the calculations of the spectral function 
of Ref.\ci{BFF89} have been employed in the calculation of corrections
to the PWIA arising from FSI between the struck nucleon and the spectator
system.
Our many-body treatment of FSI is based on much the same 
scheme as applied in I and is therefore predicated on the eikonal and 
frozen approximations.  However, the new treatment allows us
to go beyond the simple convolution expression for the response and
explicitly take into account the dependence on the initial momentum of 
the struck nucleon, which is averaged over in the CGA.

Numerical results show that the momentum dependence of the generalized 
folding function produces a sizable effect in the low-energy tail
of the $y$-scaling function in the range of momentum transfer 
1 $< |{\bf q}| <$ 2 GeV/c covered by the inclusive electron-nucleus 
scattering data analyzed in Refs.\ci{Benhar91,Benhar93,Benhar94}. 
While repeating the analysis of Refs.\ci{Benhar91,Benhar93,Benhar94}
within the approach proposed here
would certainly be of great interest, it must be pointed out 
that using a momentum-dependent generalized folding function
would in no way help to improve the agreement between the CGA and 
the data. 
As shown in I, the discrepancy between the CGA results and the measured 
cross sections
increases as $|{\bf q}|$ increases, while the suppression of FSI 
due to the momentum dependence of the folding function appears to be
larger at lower momentum transfer. A different mechanism leading 
to a quenching of FSI and exhibiting the {\it opposite} momentum-transfer 
dependence, such as the one associated with the color-transparency model 
employed in Refs.\ci{Benhar91,Benhar93,Benhar94},  
still seems to be needed to reconcile theory and data.

\section{acknowledgments}
This research was supported in part by the U. S. National Science
Foundation under Grant No.~PHY-9900713 (JWC) and by the Italian MIUR 
through the {\em Progetto di Ricerca di Interesse Nazionale: Fisica 
Teorica del Nucleo Atomico e dei Sistemi a Molti Corpi}.


\begin{figure}
\vspace*{1.in}
\centerline
{\epsfig{figure=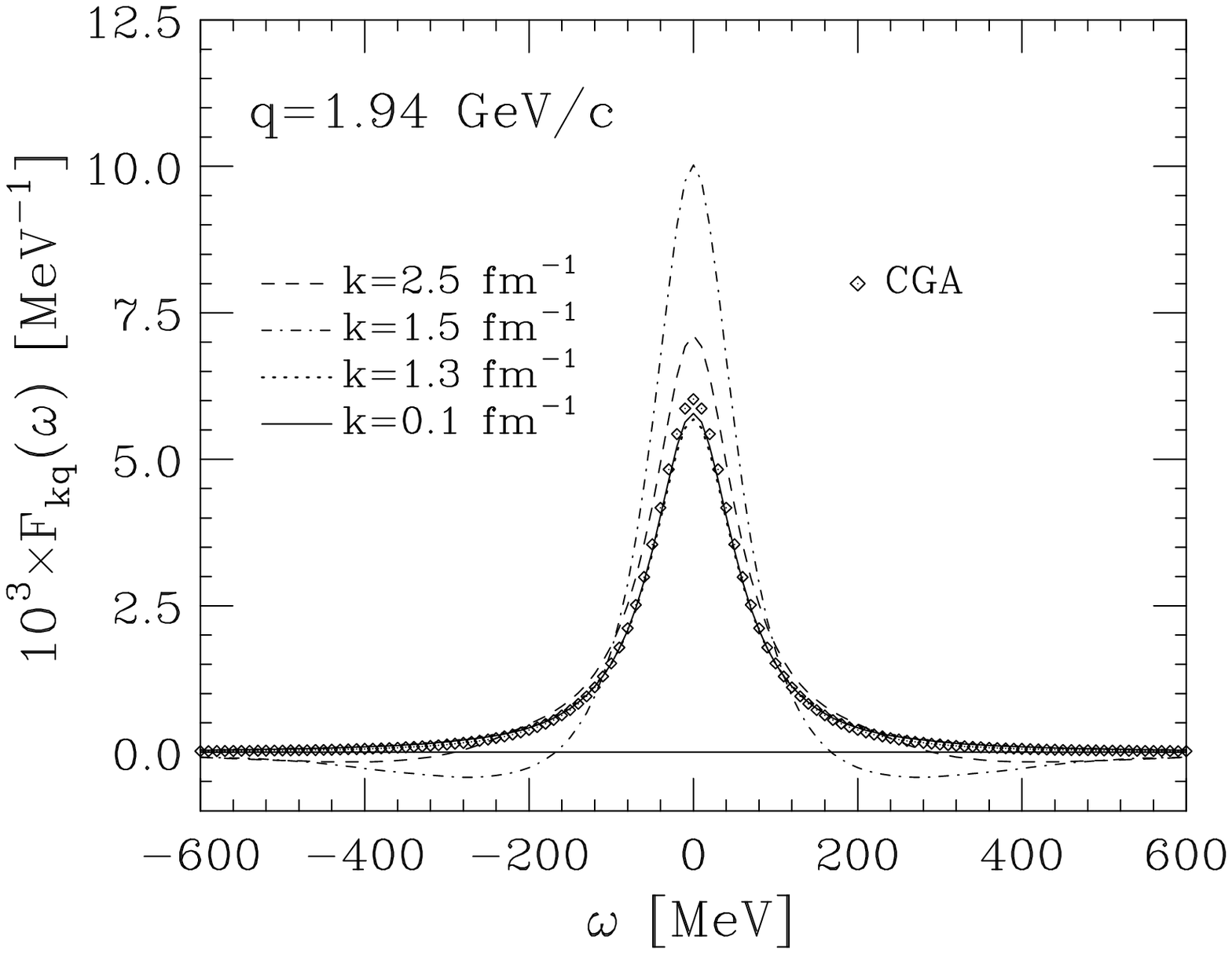}}
\vspace*{.2in}
\caption{
Dependence of the generalized folding function $F_{{\bf k},{\bf q}}(\omega)$ 
of Eqs.~(\protect\ref{eq:fol8})--(\protect\ref{eq:calE-last})
on $k=|{\bf k}|$ and $\omega$,
at $q=|{\bf q}| = 1.94$ GeV/c. The diamonds show the results of the approach 
of Ref.\protect\ci{Benhar91}, in which the dependence of the folding function
upon the initial nucleon momentum ${\bf k}$ is neglected.
}
\label{GFF1}
\end{figure}

\clearpage

\begin{figure}
\vspace*{1.in}
\centerline
{\epsfig{figure=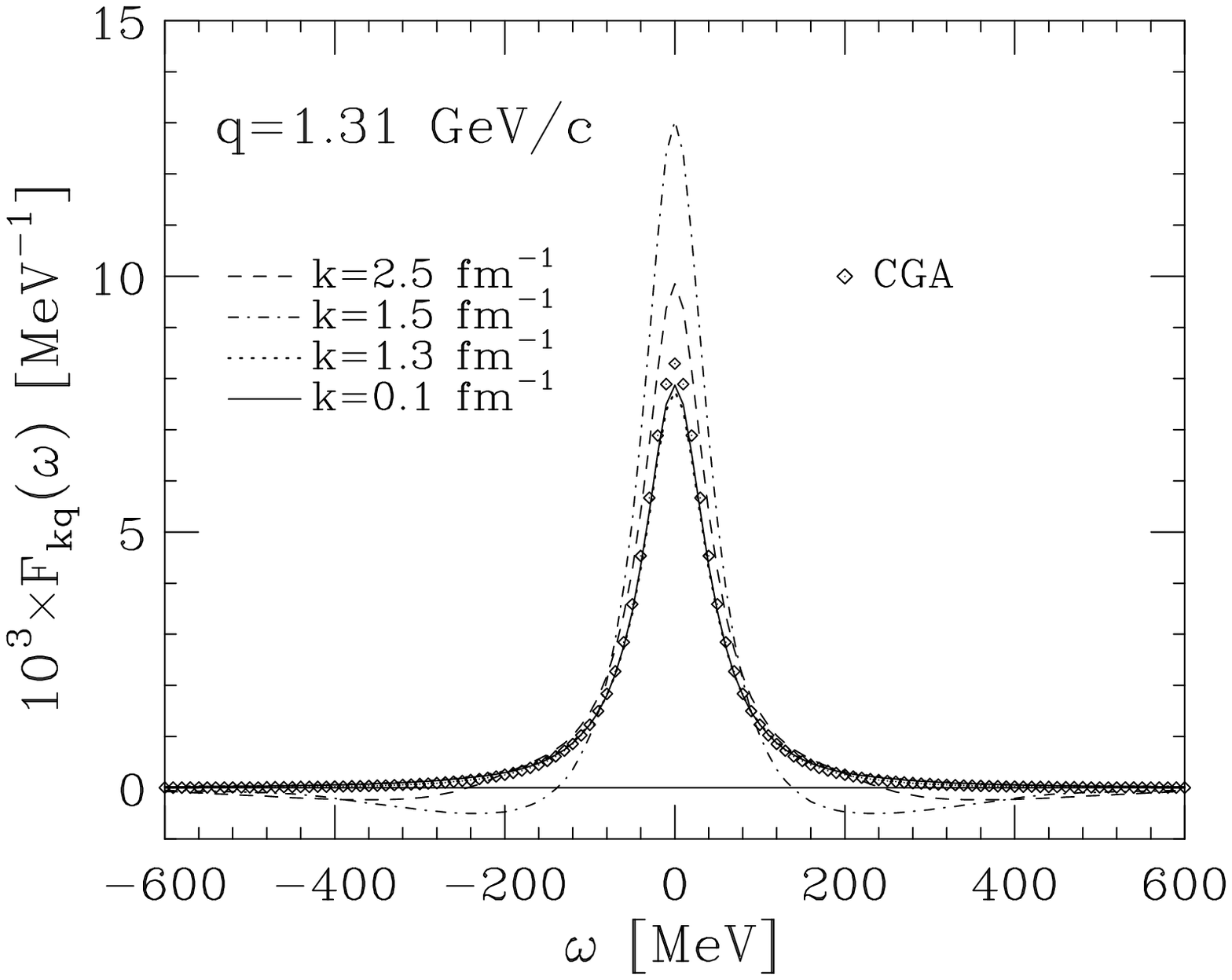}}
\vspace*{.2in}
\caption{
Same as in Fig.\protect\ref{GFF1}, but for momentum transfer
$q=|{\bf q}|$ = 1.31 GeV/c.
}
\label{GFF2}
\end{figure}

\clearpage

\begin{figure}
\vspace*{1.in}
\centerline
{\epsfig{figure=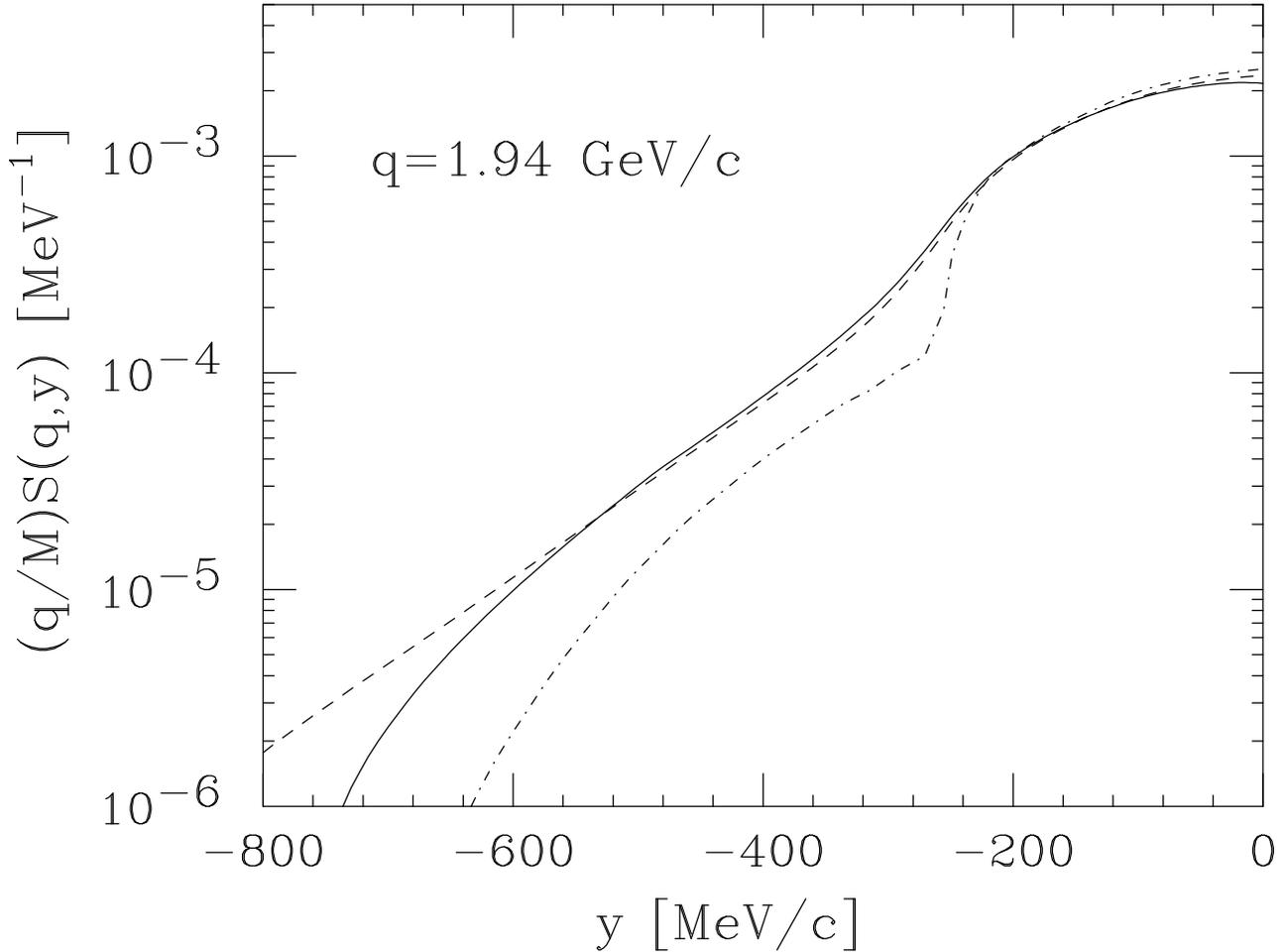}}
\vspace*{.2in}
\caption{
Dependence of the nuclear-matter scaling function $F({\bf q},y)$ defined by 
Eqs.~(\protect\ref{def:Fy})--(\protect\ref{def:y}) on the
scaling variable $y$, at
$q=|{\bf q}|$ = 1.94 GeV/c. The dash-dot line shows the PWIA result, while 
the solid and dashed lines correspond respectively to calculations
carried out using the generalized folding function of 
Eqs.~(\protect\ref{eq:fol8})--(\protect\ref{eq:calE-last}) and the 
$|{\bf k}|$-independent CGA folding function of Ref.\protect\ci{Benhar91}.
}
\label{FY1}
\end{figure}

\clearpage

\begin{figure}
\vspace*{1.in}
\centerline
{\epsfig{figure=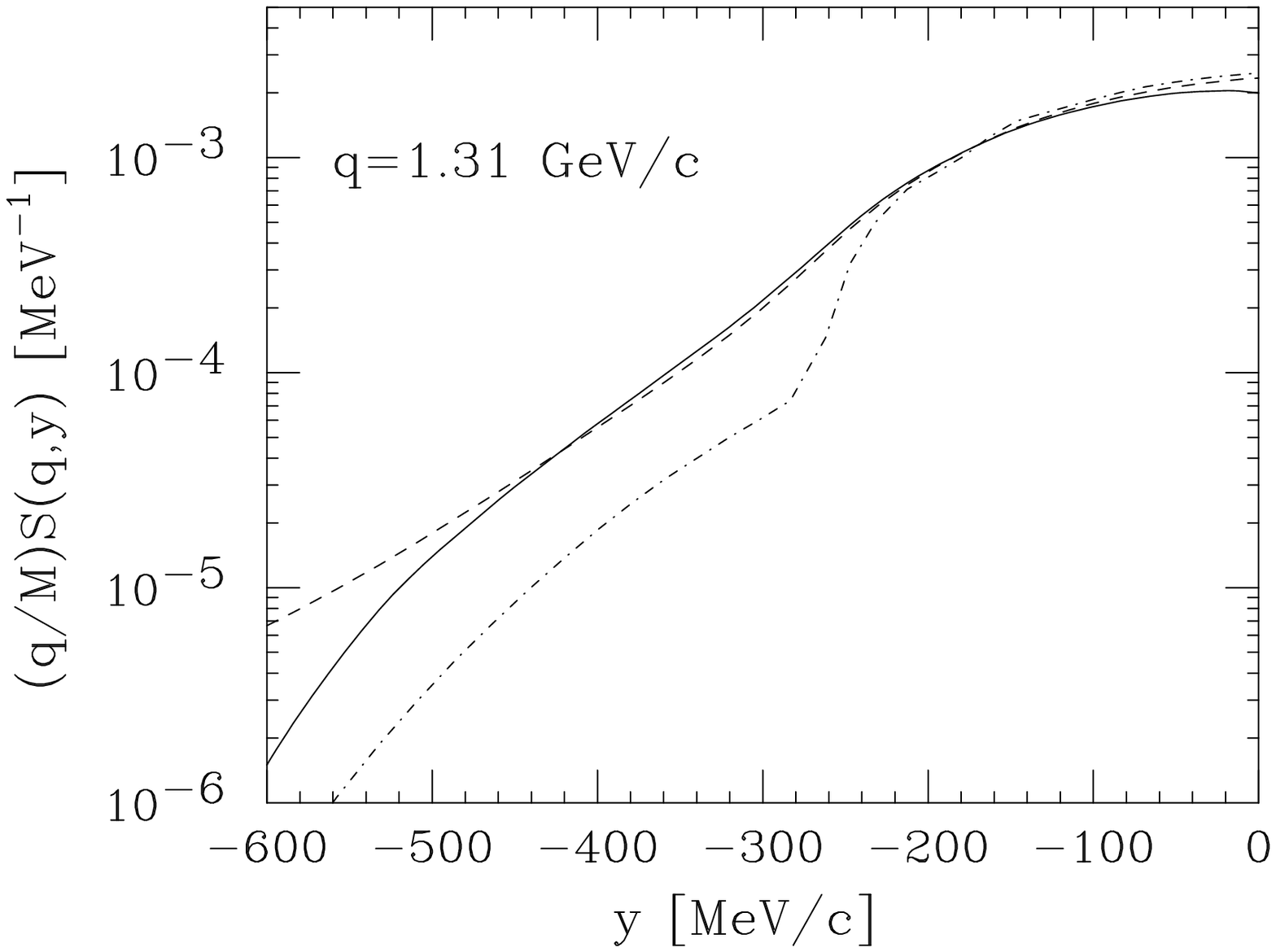}}
\vspace*{.2in}
\caption{
Same as in Fig.\protect\ref{FY1}, but for momentum transfer
$q=|{\bf q}|$ = 1.31 GeV/c.
}
\label{FY2}
\end{figure}

\end{document}